\documentclass[%
    superscriptaddress,
    preprintnumbers,
    amsmath,amssymb,
    aps,
]{revtex4-2}
\usepackage{newpxtext}
\usepackage{mathpazo}
\usepackage{amsmath,amssymb}
\usepackage{graphicx}
\usepackage[caption=false]{subfig}
\usepackage{braket}
\usepackage{booktabs}
\usepackage[bb=txof,cal=cm]{mathalpha}
\usepackage{xcolor}

\begin{document}
\preprint{KEK-TH-2526}
\preprint{J-PARC-TH-0287}
\preprint{RIKEN-iTHEMS-Report-23}
\title{Survival Probability of Unstable States in Coupled-Channels \\
-non-exponential decay of \lq\lq threshold-cusp\rq\rq~ }
\date{\today}
\author{Wren A. Yamada}
\email{wren-phys@g.ecc.u-tokyo.ac.jp}
\affiliation{Department of Physics, Faculty of Science, University of Tokyo, 7-3-1 Hongo Bunkyo-ku Tokyo 113-0033, Japan}%
\affiliation{Theory Center, Institute of Particle and Nuclear Studies (IPNS), High Energy Accelerator Research Organization (KEK), 1-1 Oho, Tsukuba, Ibaraki, 205-0801, Japan}%

\author{Osamu Morimatsu}
\email{osamu.morimatsu@kek.jp}
\affiliation{Department of Physics, Faculty of Science, University of Tokyo, 7-3-1 Hongo Bunkyo-ku Tokyo 113-0033, Japan}%
\affiliation{Theory Center, Institute of Particle and Nuclear Studies (IPNS), High Energy Accelerator Research Organization (KEK), 1-1 Oho, Tsukuba, Ibaraki, 205-0801, Japan}%

\author{Toru Sato}
\email{tsato@rcnp.osaka-u.ac.jp}
\affiliation{Research Center for Nuclear Physics (RCNP), Osaka University, Ibaraki, Osaka 567-0047, Japan}%
\author{Koichi Yazaki}
\email{koichiyzk@yahoo.co.jp}
\affiliation{RIKEN iTHEMS, Wako, Saitama 351-0198, Japan}%

\begin{abstract}
We investigate the survival probability of unstable states, the time-dependence of an initial state, in coupled channels.
First, we extend the formulation of the survival probability from single channel to coupled channels (two channels).
We derive an exact general expression of the two-channel survival probability using uniformization, a method which makes the coupled-channel S matrix single-valued, and the Mittag-Leffler expansion, i.e.\,a pole expansion.
Second, we calculate the time dependence of the two-channel survival probability by employing the derived expression.
It is the minimal distance between the pole and the physical region in the complex energy plane, not the imaginary part of the pole energy, which determines not only the energy spectrum of the Green's function but also the survival probability.
The survival probability of the \lq\lq threshold-cusp\rq\rq~caused by a pole on the unusual complex-energy Riemann sheet is shown to decay, not grow in time though the imaginary part of the pole energy is positive.
We also show that the decay of the \lq\lq threshold-cusp\rq\rq~is non-exponential.
Thus,  the \lq\lq threshold-cusp\rq\rq~is shown to be a new type of unstable mode, which is found only in coupled channels.
\end{abstract}
\maketitle
Recently, many peaks have been observed in various processes at the threshold of hadronic channels, which are potential candidates of exotic hadrons \cite{CHEN20161,ESPOSITO20171,LEBED2017143,ALI2017123,Guo:2017jvc, karliner,RevModPhys.90.015003,LIU2019237}.
These peaks are supposed to be caused by the poles of the coupled-channel $S$ matrix located on the (unusual) complex-energy Riemann sheets different from those on which the poles of usual resonances are located, or in their neighborhood \cite{PhysRevD.105.014034}.
In particular, the imaginary parts of the complex energy of these poles on the unusual sheets are positive, while those of usual resonances are negative \cite{PhysRevD.105.014034}.
Naively, the states associated with these poles on the unusual sheets grow in time in contrast to the resonant states' decay as follows.
Consider the two-channel scattering with the time-dependent Schr\"{o}dinger equation for definiteness. Time-evolution of the wave function is given as
\begin{align}
  i \frac{\partial}{\partial t} \Psi(1,2;t)=H \Psi(1,2;t), 
\end{align}
where $1$ and $2$ symbolically represent the coordinates of channels 1 and 2, respectively. For the eigenstate $\Phi(1,2)$ of the Hamiltonian with eigenenergy $E$, time-evolution of $\Psi$ is
\begin{align}
  \Psi(1,2;t) = \Phi(1,2) e^{-iEt}.
\end{align}
The complex eigenenergy of the time-independent Schr\"{o}dinger equation gives the pole of the $S$ matrix and vice versa.
Then, the probability of the state $\Psi(1,2;t)$ is
\begin{align}
  |\Psi(1,2;t)|^2 = |\Phi(1,2)|^2 e^{2{\rm Im} Et}
\end{align}
which grows (decays) in time if ${\rm Im} E > 0$ (${\rm Im} E < 0$).
Of course, this argument is too naive, and the time dependence of the unstable state has to be treated in a more legitimate manner.

There has been a history of study concerning the time-dependence of unstable states 
\cite{khalfin1958contribution,goldberger1964collision,Fonda:1978dk,nakazato1996temporal,RamirezJimenez:2018jaj,RamirezJimenez:2021pok,Chiu:1977ds,khalfin1982proton,chiu1982time,PhysRevLett.60.2246,rothe2006violation,wilkinson1997experimental,Hatano:2014cga,Ordonez:2016wyh},
which is called survival probability \cite{Fonda:1978dk,nakazato1996temporal}.
Theoretically, it has been shown that the law of resonance decay deviates from the exponential behavior at short and long times.
The time-dependence is expected to be quadratic at short times \cite{Chiu:1977ds,RamirezJimenez:2018jaj} and inverse power at long times \cite{Fonda:1978dk,RamirezJimenez:2018jaj}.
Possible implications of quadratic short-time behavior in the Zeno effects and the proton decay have been discussed in Ref.\,\cite{Chiu:1977ds,chiu1982time,khalfin1982proton}.
Experimentally, in nuclear and particle physics, there have been attempts to observe deviations from purely exponential behavior, but they have been unsuccessful
(see Ref.\,\cite{PhysRevLett.60.2246} and references therein).
The non-exponential behavior at large times was found in an experiment measuring the
luminescence decays of dissolved organic materials after pulsed laser excitation \cite{rothe2006violation}.
It should be noted, however, that the study of  the time-dependence of the unstable state has been limited to the problems of one open channel, single-channel 
as far as we know, while the time dependence of the unstable state associated with the \lq\lq threshold-cusp\rq\rq~is a problem in coupled channels.

The purpose of the present letter is two-fold.
The first is to extend the formulation of the survival probability in single channels to the one in coupled channels (two channels).
We derive an exact general expression of the two-channel survival probability from the Mittag-Leffler expansion \cite{arfken2013mathematical,Nussenzveig:1972tcd}, i.e.\,a pole expansion, of the Green's function in the uniformization variable \cite{NEWTON195829,KATO1965130}, in terms of which the Green's function is represented single-valued.
This is based on the method developed and applied to the study of unstable states in coupled channels in Refs.\,\cite{RAMIREZJIMENEZ201818,PhysRevC.102.055201,Yamada:2021cjo,PhysRevLett.129.192001,PhysRevD.105.014034}.
The second is to examine the time dependence of the two-channel survival probability, particularly for the \lq\lq threshold-cusp\rq\rq~ by employing the derived expression.
We show that the survival probability for the \lq\lq threshold-cusp\rq\rq~decays, not grow in time against the above naive argument.
In general, and in particular, for the \lq\lq threshold-cusp\rq\rq~ it is the minimal distance between the pole and the physical region in the complex energy plane, not the imaginary part of the pole energy, which determines not only the energy spectrum of the Green's function but also the time dependence.
We also show that the decay of the \lq\lq threshold-cusp\rq\rq~is non-exponential in contrast to the exponential decay of the resonance.
\par
Consider an unstable state, $\ket{\psi(t)}$, time-evolving under the Hamiltonain, $H$ as, $\ket{\psi(t)}=e^{-iHt}\ket{\psi(0)}$.
Following Refs.\,\cite{Hatano:2014cga,Ordonez:2016wyh}, we define the \lq\lq survival amplitude", $\mathcal{A}(t)$, of the state $\ket{\psi(0)}$ by
\begin{align}
  \mathcal{A}(t) =  \langle \psi(0) |e^{-iHt} | \psi(0) \rangle = \frac{1}{2\pi i} \int_C de \langle \psi(0) | \frac{1}{e-H} | \psi(0) \rangle e^{-iet} = \sum_B |\braket{\psi(0)|\phi_B}|^2 e^{-ie_Bt} + \mathcal{A}_{cut}(t),
\end{align}
where
\begin{align}
  \mathcal{A}_{cut}(t) &= \frac{1}{2\pi i} \left[ \int_\infty^{0} de \langle \psi(0) | \frac{1}{e-H+i0} | \psi(0) \rangle e^{-iet} + \int_{0}^\infty de \langle \psi(0) | \frac{1}{e-H-i0} | \psi(0) \rangle e^{-iet} \right] \nonumber\\
  &= - \frac{1}{2\pi i} \int_{0}^\infty de \left[{\cal G}(e+i0) - {\cal G}(e-i0)\right] e^{-iet}.
\end{align}
$e$ is the dimensionless energy with $e=0$ taken as the lowest threshold and $e=1$ as the other threshold in the two-channel case. $\{\ket{\phi_B}\}$ and $\{e_B\}$ are the bound states and their eigenenergies, respectively. The sum is taken over all bound states.
$C$ is the contour from infinity to the lowest eigenenergy along the real axis below the cut and back to infinity above the cut. 
$\mathcal{G}(e)$ is the Green's function of the Hamiltonian.
The probability that $\ket{\psi(t)}$ is in its initial state, which we will call the survival probability, $P(t)$, is then given by $P(t)=|\mathcal{A}(t)|^2$.
\par
The Green's function, $\mathcal{G}(e)$, can be expanded by a Mittag-Leffler expansion \cite{arfken2013mathematical,Nussenzveig:1972tcd}  in the uniformization variable \cite{NEWTON195829,KATO1965130}, symbolically written as $u$ for the moment, as
\begin{align}
  {\cal G}(e)
   =\sum_n r_n\mathcal{G}_n, \qquad
  {\cal G}_n
   &= \frac{1}{u-u_n}.
\end{align}
The residues, $r_n$, satisfy the condition, $\sum_n r_n = 0$.
Then, the cut contribution, $\mathcal{A}_{cut}$, can be expressed by a sum of pole terms as \cite{RAMIREZJIMENEZ201818,Ordonez:2016wyh}
\begin{align}
  \mathcal{A}_{cut}(t) = \sum_n r_n \mathcal{A}_n, \qquad
  \mathcal{A}_n &= - \frac{1}{2\pi i} \int_{0}^\infty de \left[{\cal G}_n(e+i0) - {\cal G}_n(e-i0)\right] e^{-iet}.
  \label{eq:sv_mle}
\end{align}
\par
Before proceeding to the two-channel survival probability, we briefly review the single-channel survival probability for comparison.
\par
For the single-channel case, dimensionless momentum, $k =  e^{1/2}$, plays a role of the uniformization variable.
We can define the minimal distance between the pole and the physical domain on the $e$-plane, $d_n^e$, based on the distance on the uniformized $k$-plane, $d_n^k$, as,
\begin{align}
  d_n^e  = \left|\frac{de}{dk}\right|_{k=k_{n}}\times d_n^k = 2|k_n {\rm Im} \, k_n|,
\end{align}
where $k_n$ is the position of the pole of the Green's function on the $k$-plane. Pole energy $e_n$ is given by $e_n=k_n^2$.
For resonance poles sufficiently distant from the threshold, the distance, $d_n^e$, coincides with the imaginary part of the pole energy, i.e., $d_n^e \approx |{\rm Im} \, e_n| = \Gamma_n/2$, where $e_n=\varepsilon_n-i\Gamma_n/2$.
\par
The pole contribution of the Green's function, $\mathcal{G}_n$, is given as
\begin{align}
  {\cal G}_n = \frac{1}{k - k_{n}} 
  = \frac{e^{1/2} + e_{n}^{1/2}}{e - e_{n}} ,
\end{align}
and the survival probability, $\mathcal{A}_n$, as
\begin{align}\label{eq:single_sv}
  \mathcal{A}_n = -\frac{1}{i\pi} \int_0^\infty de \frac{\sqrt{e}} {e - e_n} e^{-iet} = -\frac{1}{i\pi} \left[ \sqrt{\frac{\pi}{it}} - i \pi \sqrt{e_n} \, e^{-i e_n t} {\textrm{erfc}}(i\sqrt{i e_n t}) \right]  \equiv I(t,e_n).
\end{align}
erfc$(z)$ is the complementary error function.
(Note that $\sqrt{t}I(t,e_n)$ is a function of $te_n$ implying a time-scale of $|e_n|^{-1}$.)
\par
Consider the contributions from a typical resonance pole ($e_n=\varepsilon_n-i\Gamma_n/2$, $\Gamma_n/\varepsilon_n\ll 1$)  and its counterpart, anti-resonance pole, ($e_n^\ast=\varepsilon_n+i\Gamma_n/2$).
\par
In the vicinity of the resonance pole energy, ${\cal G}_n$ can be approximated as
\begin{align}
  {\cal G}_n = \frac{1}{k - k_{n}} 
  \approx \frac{2 e_n^{1/2}}{e - e_n} 
  = \frac{r^e_n}{e - e_n},
\end{align}
where $r^e_n=2 e_n^{1/2}$.
$|{\cal G}_n|^2$ is then given by the Breit-Wigner form \cite{Breit:1936zzb},
\begin{align}
  |{\cal G}_n|^2 \approx \left|\frac{r^e_n}{e - e_n}\right|^2 = \frac{|r^e_n|^2}{(e-\varepsilon_n)^2 + \Gamma_n^2/4}.
\end{align}
The width of the resonance peak in the spectrum is $\Gamma_n\approx 2 d_n^e$.
\par
The survival amplitude can also be approximated for a resonance and an anti-resonance as,
\begin{align}
  I(t,e_n) &\approx  -\frac{r^e_n}{2\pi i} \int_{-\infty}^\infty de \frac{1} {e - e_n} e^{-iet}
  =
   - \theta(t) r^e_n e^{-i e_n t},
\\
  I(t,e_n^\ast) &\approx  -\frac{r^{e\ast}_n}{2\pi i} \int_{-\infty}^\infty de \frac{1} {e - e_n^\ast} e^{-iet}
  =
 \theta(-t) r^{e\ast}_n e^{-i e_n^\ast t}.
\end{align}
where $\theta(t)$ is the Heaviside step function.
The survival probability for the resonance and anti-resonance contributions are, therefore,
\begin{gather}
  |I(t,e_n)|^2 \approx 
  \theta(t) |r^e_n|^2 e^{-\Gamma_n t},
  \quad\quad\quad
  |I(t,e_n^\ast)|^2 \approx 
  \theta(-t) |r^e_n|^2 e^{\Gamma_n t},
\end{gather}
respectively.
When contributions from other poles are negligible, one expects an exponential decay (growth) of the survival probability in the $t>0$ ($t<0$) region with a characteristic time scale, which in this case is the \lq\lq lifetime," $1/\Gamma_n$. 
This characteristic time scale is the inverse of the width of the spectrum, which corresponds to the minimal distance between the pole and the physical region.
\par
At long and short times, the above approximation breaks down, resulting in deviations from the exponential behavior of the survival probability.
The large $t$ behavior of $I(t,e_n)$ can be obtained from the asymptotic expansion of the complementary error function in Eq.\,\eqref{eq:single_sv} as
\begin{align}
  I(t,e_n) &= \frac{i}{\sqrt{4\pi}}e_n^{-1} e^{i\pi/4} e^{-it} t^{-3/2} + O(t^{-5/2}),
\end{align}
which coincides with the $t^{-3}$ inverse-power decay law of the survival probability obtained in previous studies \cite{RamirezJimenez:2018jaj}. 
At short times (small $t$), the survival amplitude, $\mathcal{A}(t)$, is known to behave as quadratic in time as long as all the matrix elements $\langle \psi(0)| H^n |\psi(0) \rangle$ are finite \cite{RamirezJimenez:2021pok}.
As can be seen from Eq.\,\eqref{eq:single_sv}, however, each pole contribution, $I(t,e_n)$, has a singular term, $t^{-1/2}$.
By summing up contributions from all poles, the leading singular terms of $t^{-1/2}$ cancel out due to the condition of the residues, $\sum_n r_n = 0$.
Other nonleading singular terms should also vanish so that $\mathcal{A}(t)$ is quadratic.
Nevertheless, to discuss the short-time behavior of the survival amplitude, $\mathcal{A}(t)$, all the pole contributions must be summed up, which will not be pursued here.
\par
\begin{figure}[htpb]
  \includegraphics[width=\linewidth]{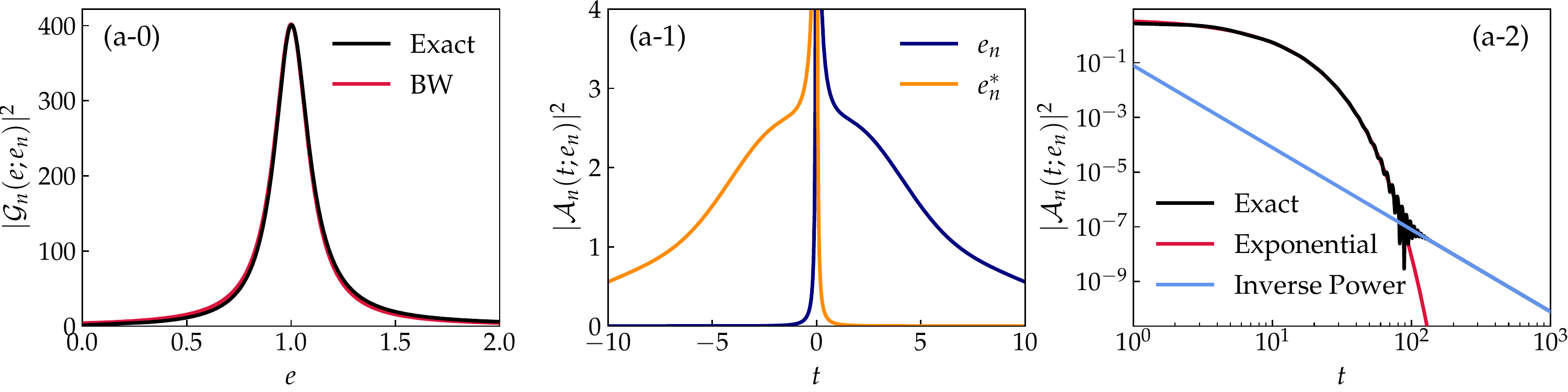}
\caption{Energy spectrum $|\mathcal{G}_n|^2$ (a-0), and the survival probability $|\mathcal{A}_n|^2$ (a-1,a-2), of a typical resonance and anti-resonance.  $k_n=1.00125 - 0.049937 i$, $e_n=1.0-0.1 i$, and $d^e_n=0.100124$.
 Labels, \lq\lq Exact\rq\rq, \lq\lq Exponential\rq\rq~and \lq\lq Inverse Power\rq\rq, in (a-2), correspond to $\mathcal{A}_n$ of Eqs.\,(10), (15) and (16), respectively.
}
\label{fig:single_chan}
\end{figure}

The energy spectrum $|\mathcal{G}_n|^2$, and the survival probability $|\mathcal{A}_n|^2$, of a typical resonance and anti-resonance are shown in Fig.\,\ref{fig:single_chan}.
We can observe a transition from an exponential decay with a lifetime $\Gamma_n^{-1}$ to an inverse-power decay with a power of $t^{-3}$, coinciding with the above argument.
We can also observe an oscillation of the survival probability in the transition region.
\par
\begin{figure}[htbp]
      \includegraphics[width=0.33\linewidth]{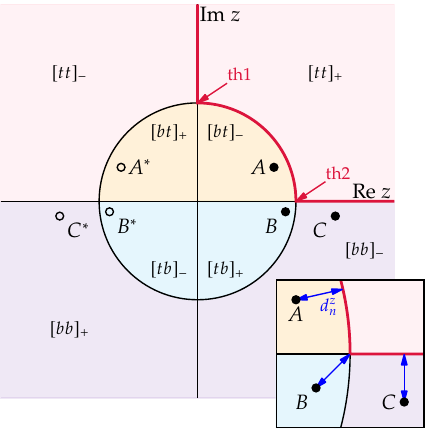}
      \caption{Schematic figure of the uniformized $z$-plane. The red line shows the physical region. th1 (th2) is the lower (upper) threshold.
      $A$ ($A^\ast$), $B$ ($B^\ast$), and $C$ ($C^\ast$) correspond to the poles specified in Tab.\,1.}
      \label{fig:unif_2}
\end{figure}

Let us extend the argument above to the two-channel case.
The uniformization variable, $z$, of a two-channel system can be given by \cite{NEWTON195829,KATO1965130},
\begin{gather}
  z =  k_1 + k_2 = e^{1/2} + (e-1)^{1/2}, \\
  k_1 = \frac{1}{2} \left( z + \frac{1}{z} \right),  \quad k_2 = \frac{1}{2} \left( z - \frac{1}{z} \right), \quad
  e = k_1^2 = k_2^2 + 1 = \frac{1}{4} \left( z^2 + \frac{1}{z^2} +2 \right) ,
\end{gather}
where $e$ is the dimensionless energy, and $k_i$ is the momentum of channel $i$ ($i=1,2$). 
Let us specify (the) eight domains of the four-sheeted $e$-plane by a set of the sign of the imaginary part of channel momenta, together with a label $+$ ($-$) representing the upper half (lower half) part of the $e$-plane. e.g.\ [$tb$]$_+$ means ${\rm Im}\, k_1 > 0$, ${\rm Im}\, k_2 < 0$, and ${\rm Im}~e >0$. 
The correspondence between the $z$-plane and the $e$-plane is shown in Fig.\,\ref{fig:unif_2}.
\par
Analogously to the single-channel case, we can define the minimal distance between the pole and the physical domain on the $e$-plane, $d_n^e$, based on the distance on the uniformized $z$-plane, $d_n^z$, as,

\begin{align}
  d_n^e = \left|\frac{de}{dz}\right|_{z=z_n}\times d_n^z = \frac{1}{2}\left| z - \frac{1}{z^3} \right|_{z=z_n} \times d_n^z,
\end{align}
where,
\begin{align}
  d_n^z = 
  \begin{cases}
  1 - |z_n| & \text{$[bt]_-$ pole}\\
  |1 - z_n | & \text{$[tb]_+$ pole}\\
  {\rm Im} \, z_n & \text{$[bb]_-$ pole}
  \end{cases}.
\end{align}
$z_n$ is the pole position of Green's function on the $z$-plane. Pole energy $e_n$ is given by $e_n=(z_n+z_n^{-1})^2/4$. 
For resonance poles on $[bt]_-$, and $[bb]_-$, sufficiently distant from the threshold, $d_n^e \approx |{\rm Im} \, e_n| = \Gamma_n/2$, where $e_n=\varepsilon_n-i\Gamma_n/2$.
In contrast, for poles near the upper threshold, especially the ones on $[tb]_+$, which cause enhanced \lq\lq threshold cusp\rq\rq structures in the spectra, $d_n^e \approx 2\left|e_n-1\right|$.
\par
The pole contribution of the Green's function, $\mathcal{G}_n$, is explicitly given as
\begin{align}\label{G_n_exact}
  {\cal G}_n
   &= \frac{1}{z-z_n} = \frac{\left(1-\frac{1}{z_n^2}\right) e^{1/2} + \left( 1 +  \frac{1}{z_n^2} \right) (e-1)^{1/2} + \frac{2}{z_n} e^{1/2}(e-1)^{1/2} + z_n - \frac{1}{z_n} (2e-1) }
  {4(e - e_n)},
\end{align}
and the survival probability, $\mathcal{A}_n$, as
\begin{align}\label{A_n_exact}
  \mathcal{A}_n =& -\frac{1}{4i\pi}
  \Biggl[ \left(1-\frac{1}{z_n^2}\right) \int_0^\infty de \frac{\sqrt{e}} {e - e_n} e^{-iet}
  + \left( 1 +  \frac{1}{z_n^2} \right) \int_1^\infty de \frac{\sqrt{e-1}} {e - e_n} e^{-iet}
  +i \frac{2}{z_n} \int_0^1 de  \frac{\sqrt{e(1-e)}} {e - e_n} e^{-iet}
\Biggr] \nonumber\\
 =& \frac{1}{4i\pi} \left[ \left(1-\frac{1}{z_n^2}\right)i\pi I(t,e_n) + \left( 1 +  \frac{1}{z_n^2} \right) e^{-it} i\pi I(t,e_n-1) - i \frac{2}{z_n} J(t,e_n) \right],
\end{align}
where $I(t,e_n)$ is given by Eq.\,\eqref{eq:single_sv} and
\begin{gather}
  J(t,e_n) = \int_0^1 de  \frac{\sqrt{e(1-e)}} {e - e_n} e^{-iet}.\label{eq:ij}
\end{gather}
From Eqs.\,\eqref{eq:sv_mle}, \eqref{A_n_exact}, \eqref{eq:single_sv} and \eqref{eq:ij}, we obtain an exact general expression of the two-channel survival amplitude.
$I(t,e_n)$ corresponds to the pole contribution in the single-channel case.
The pole contribution in the two-channel case, $\mathcal{A}_n(t,z_n)$, comprises three terms. 
Two terms with $I(t,e_n)$ and $I(t,e_n-1)$ are the single-channel contributions from the lower and upper threshold, respectively, and the additional term with $J(t,e_n)$ is a term involving both thresholds.
An analytic expression of $J(t,e_n)$ is yet to be known but can be easily calculated numerically. 
\par
Let us consider the contribution from a single pole in more detail.
\par
In the case where the pole is on $[bt]_-$ or $[bb]_-$, and is sufficiently distant from both thresholds, ${\cal G}_n$ can be approximated as,
\begin{align}\label{G_n_BW}
  {\cal G}_n = \frac{1}{z - z_{n}} 
  \approx \frac{z_n\left(1-z_n^{-4}\right)}{2} \frac{1}{e - e_n} 
  = \frac{r_n^e}{e - e_n},
\end{align}
where $r_n^e = z_n \left(1-z_n^{-4}\right)/2$. The Breit-Wigner form approximates the spectrum; hence, the survival amplitude would behave like the single-channel case.
\par
If the pole is positioned near the upper threshold, i.e., $|z_n-1|\ll 1$, ${\cal G}_n$ can be approximated as,
\begin{align}\label{G_n_DD}
  {\cal G}_n = \frac{1}{z - z_{n}} 
  \approx \frac{1+1/z_n^2}{2}\frac{1}{k_2 - k_{2n}} 
  = \frac{r_n^k}{k_2 - k_{2n}},
\end{align}
where $k_{2n}=(z_n-z_n^{-1})/2$ and $r_n^k = \left(1+z_n^{-2}\right)/2$.
$\left|{\cal G}_n\right|^2$ is then given by the Dalitz-Deloff form \cite{Dalitz:1982tb,Dalitz:1991sq} as 
\begin{align}
  \left|{\cal G}_n\right|^2 &\approx \left| \frac{r^k_n}{k_2 - k_{2n}} \right|^2 
  =
  \begin{cases}
 \displaystyle{ \frac{|r^k_n|^2} {|k_{2n}|^2 } \frac{1} {\left(1 + k \alpha\right)^2 + (k\beta)^2 } = \frac{|r^k_n|^2} {|k_{2n}|^2 } \left[1- 2 k \alpha + O\left(k^2\right) \right] } & (e > 1)
\\
  \displaystyle{ \frac{|r^k_n|^2} {|k_{2n}|^2 } \frac{1} {\left(1 + \kappa \beta\right)^2 + (\kappa \alpha)^2 } =  \frac{|r^k_n|^2} {|k_{2n}|^2 } \left[1- 2 \kappa \beta + O\left(\kappa^2\right) \right] } & (e < 1)
  \end{cases}.
\end{align}
$\alpha$ and $\beta$ are the real and imaginary parts of the complex scattering length, $a$, satisfying, $a = \alpha + i \beta  \approx -i/k_{2n}$.
The width of the peak structure in the spectrum would be $4|k_{2n}|^2=4|e_n-1|\approx2d_n^e$, rather than the imaginary part of the pole position.
\par
\begin{figure}[!htb]
  \includegraphics[width=0.5\linewidth]{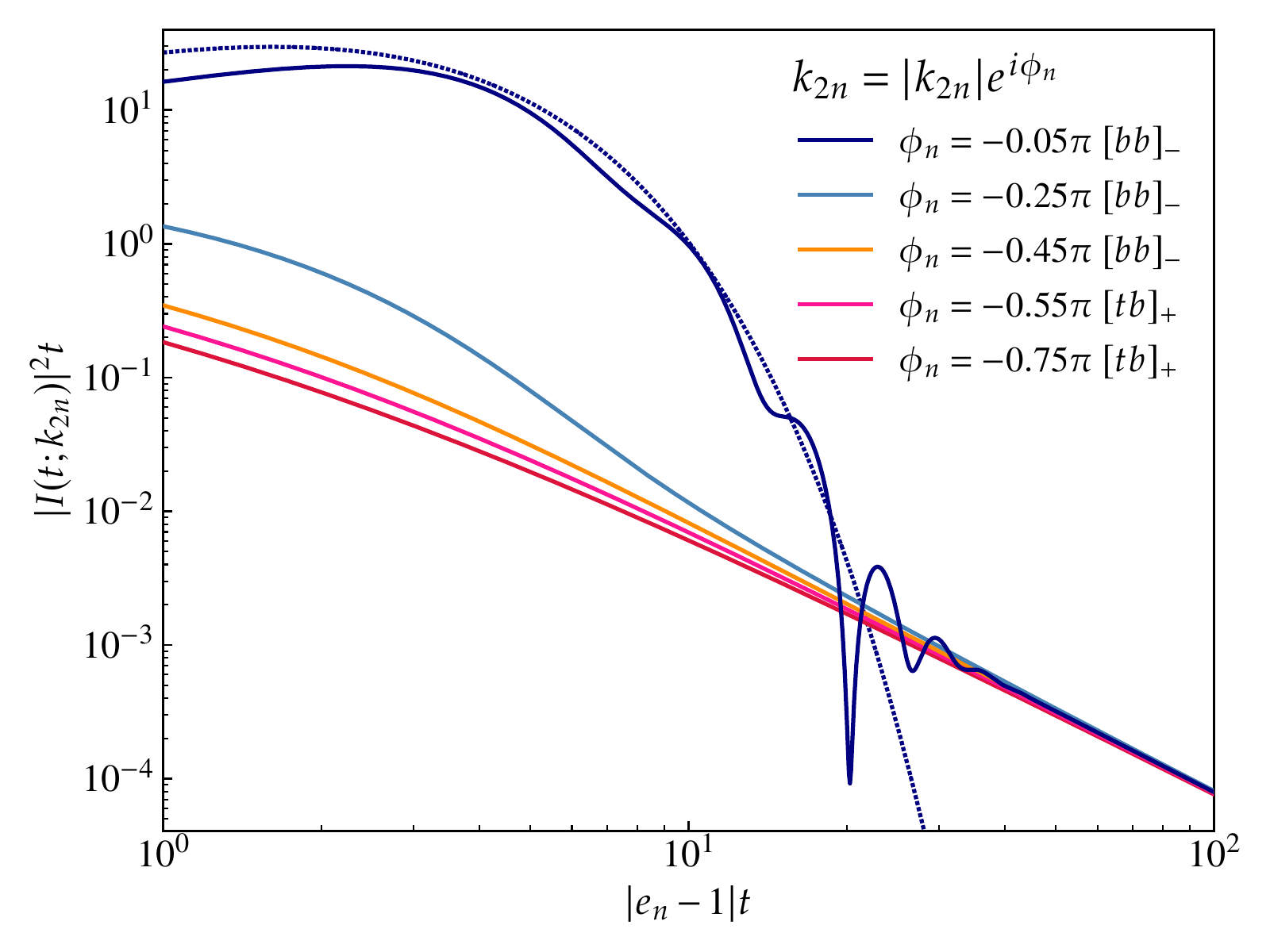}
  \caption{Argument dependence of $|I(t,k_{2n}^2)|^2$ for poles in the vicinity of the upper threshold. For reference, the blue dotted line represents the exact exponential decay of the pole of $\phi_n=-0.05\pi$.}
  \label{fig:arg_dept}
\end{figure}

The pole contribution, $\mathcal{A}_n$, can also be approximated by neglecting the lower threshold as
\begin{align}\label{A_n_single}
  \mathcal{A}_n \approx \frac{r^k_n}{2\pi i} \left( \int_{i\infty}^0 + \int_0^\infty \right) dk_2 2k_2 \frac{1}
  {k_2 - k_{2n}} e^{-i(1+k_2^2)t} 
  =
\theta(t) \displaystyle{\frac{r^k_n}{i\pi} e^{-it} I(t,k_{2n}^2)}
\end{align}
Thus,
\begin{align}
  \left|\mathcal{A}_n\right|^2 &\approx
  \theta(t)\displaystyle{ \frac{1}{\pi^2} |r^k_n|^2 |I(t,k_{2n}^2) |^2 }
\end{align}
When contributions from other poles are negligible, we expect a decay of the survival probability in the $t>0$ region governed by the behavior of $|I(t,k_{2n}^2)|^2$.
From the expression of $I$, we see a characteristic time scale of decay proportional to $|e_n-1|^{-1}$, corresponding to the inverse of the width of the peak structure in the spectrum.
The argument dependence of $|I(t,k_{2n}^2)|^2$ is shown in Fig.\,\ref{fig:arg_dept}.
Most importantly, we observe a decay rather than a growth for poles on $[tb]_+$, though they have a positive imaginary part in complex energy.
$|I(t,k_{2n})|^2$ shows a non-exponential decay for all $t$, in regions of $[tb]_+$, and in the neighboring regions on $[bt]_-$, and $[bb]_-$. 
An exponential behavior gradually emerges as the pole continuously moves away from the upper threshold but stays close to the physical region.
\par
The behavior of the survival probability at large $t$ exhibits an inverse-power decay.
The asymptotic expansion of $\mathcal{A}_n$ can be obtained by applying the method of steepest descent to $J$, leading to
\begin{align}\label{A_n_asymptotic}
  \mathcal{A}_n &\sim \frac{i}{\sqrt{4\pi}} e^{i\pi/4} e^{-it/2} t^{-3/2} \left[ e^{it/2} \left(z_n-i\right)^{-2} + e^{-it/2} \left(z_n-1\right)^{-2} \right] + O(t^{-5/2}).
\end{align}
As in the single-channel case, the survival probability decreases by a power of $t^{-3}$. 
However, there are additional oscillations due to two terms weighted by the inverse square of the distance between the pole and the two thresholds.
\par
\begin{figure}[htpb]
  \includegraphics[width=\linewidth]{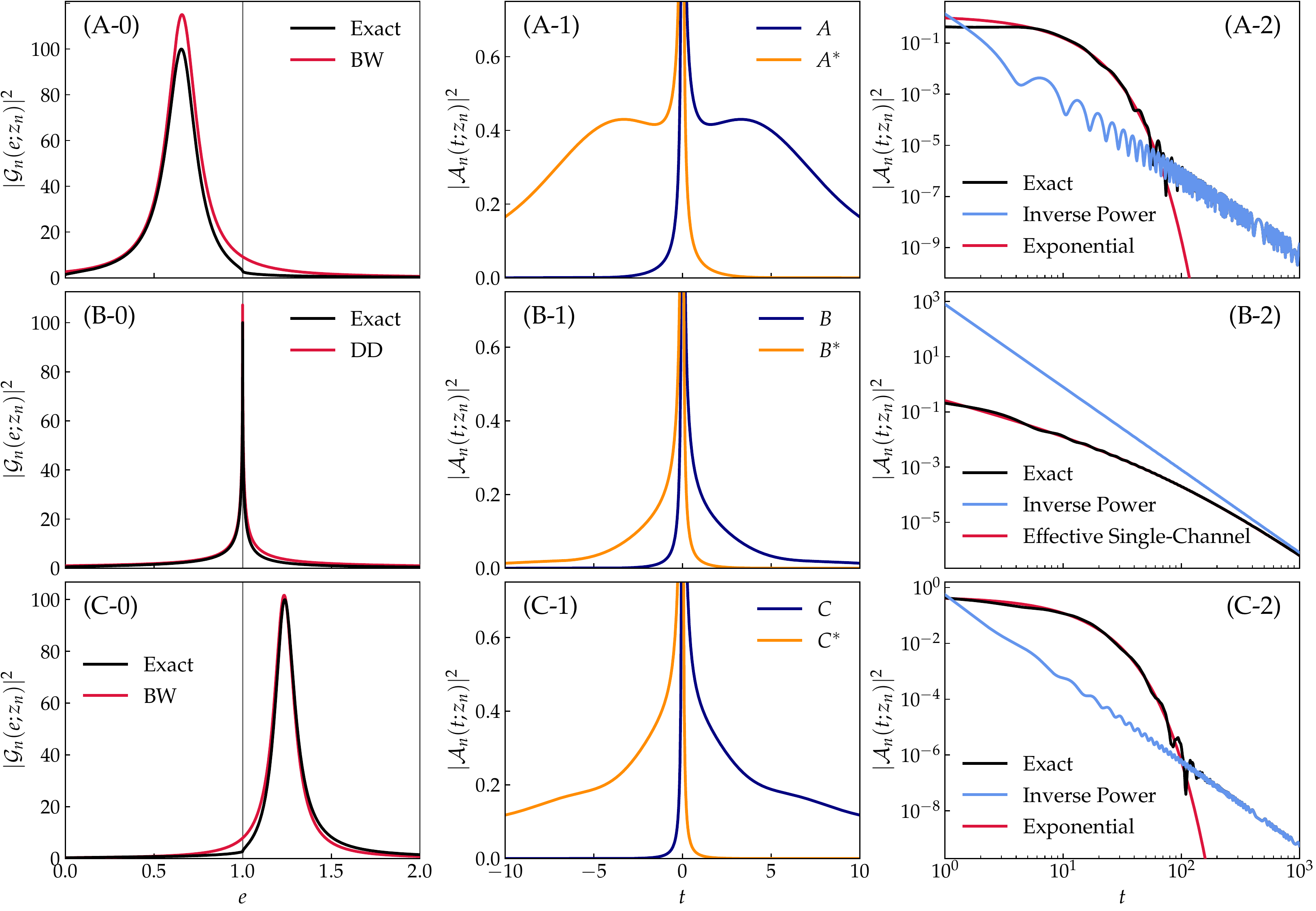}
\caption{Energy spectrum $|\mathcal{G}_n|^2$ (left), and the survival probability$|\mathcal{A}_n|^2$ (middle, right), of cases A, B, and C.
Labels, \lq\lq Exact\rq\rq, \lq\lq BW\rq\rq and \lq\lq DD\rq\rq in the left figures, correspond to $\mathcal{G}_n$ of Eqs.\,(\ref{G_n_exact}), (\ref{G_n_BW}) and (\ref{G_n_DD}), respectively.
Labels, \lq\lq Exact\rq\rq, \lq\lq Inverse Power\rq\rq and \lq\lq Effective Single-Channel\rq\rq in the right figures, correspond to $\mathcal{A}_n$ of Eqs.\,(\ref{A_n_exact}), (\ref{A_n_asymptotic}) and (\ref{A_n_single}), respectively.
}
\label{fig:main}
\end{figure}

To confirm the argument above, we consider three cases, A, B, and C, which correspond to a lower-channel resonance, a threshold cusp, and an upper-channel resonance.
Tab.\,1 shows the pole position in terms of the uniformization variable, $z_n$, its complex energy, $e_n$, the minimal distance between the pole and the physical region on the $e$-plane, $d_n^e$, and the sheet configuration where the poles are located.
Fig.\,\ref{fig:main} shows the energy spectrum, $|\mathcal{G}_n|^2$, and the survival probability, $|\mathcal{A}_n|^2$, of the three cases, A, B, and C.
\begin{table}[h]
\begin{tabular}{cccc}
\hline
\hline
&$A$&$B$&$C$\\
\hline
$z_n$&$0.7281+0.5290 i$&$0.9293-0.0707 i$&$1.6-0.1 i$\\
$e_n$&$0.6580-0.1009 i$&$0.9992+0.0107 i$&$1.234-0.0679 i$\\
$d^e_n$ & $0.1083$ & $0.0222$ & $0.0685$ \\
sheet&$[bt]_-$&$[tb]_+$&$[bb]_-$\\
\hline
\end{tabular}
\caption{Pole position in terms of the uniformization variable, $z_n$, its complex energy, $e_n$, the minimal distance between the pole and the physical region, $d_n^e$, and the sheet configurations of where the poles are located for cases A, B, and C.}
\label{table_1}
\end{table}

As we can see, the results in Fig.\,\ref{fig:main} coincide with our arguments above.
For resonance cases A and C, one can see a transition from exponential to inverse-power decay. In contrast, the \lq\lq threshold cusp\rq\rq case, B, exhibits a non-exponential decay for all time regions matching the behavior of $|I|^2$.
The width, $\Gamma$, of the spectrum's peak for cases A and C are approximately $\Gamma\approx2|\textrm{Im}~e_n|\approx 2d_n^e$, whereas $\Gamma\approx4|e_n-1|\approx 2d_n^e$ for case B. 
In all cases, the spectra' width corresponds with the minimal distance between the pole and the physical region.
The time scale for the exponential decay in cases A and C are given by $|\textrm{Im}~e_n|^{-1}$, and for case B, is given by $|e_n-1|$, which in both cases correspond to the inverse of the width of the spectrum.
We also point out that for cases A and C, the resonance (anti-resonance) contribution, which is the contribution from the pole with a negative (positive) imaginary part of complex energy, is dominant in the $t>0$ ($t<0$) region. 
On the contrary, for case B, the contribution from the pole with a positive (negative) imaginary part of complex energy is dominant in the region $t>0$ ($t<0$).
\par
\begin{figure}[htpb]
  \includegraphics[width=\linewidth]{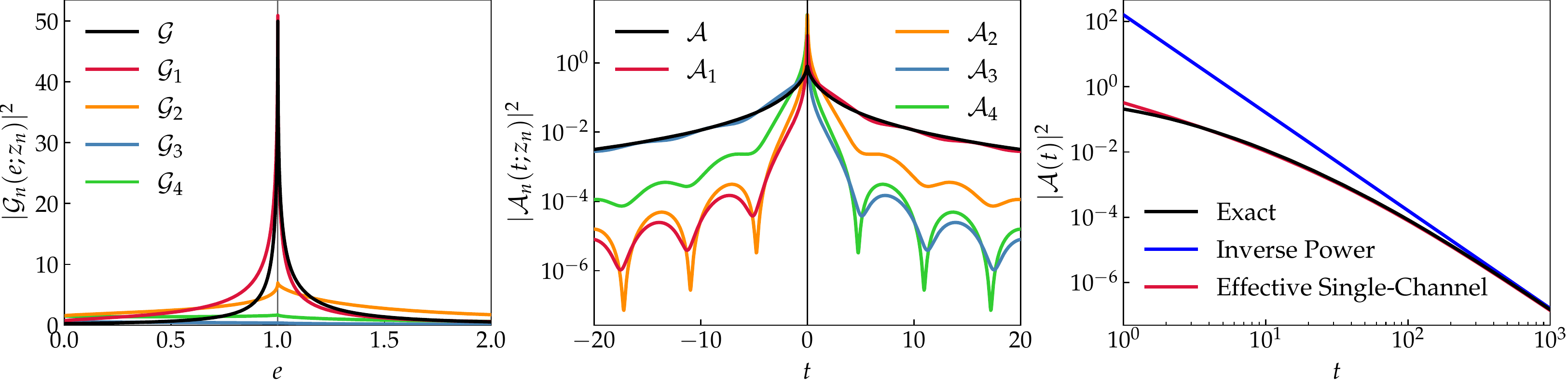}
\caption{Energy spectrum (left) and the survival probability (center, right) for $m = 1.1$, $\gamma_1=0.1$, $\gamma_2=0.9$.
\lq\lq Exact\rq\rq, \lq\lq Inverse Power\rq\rq and \lq\lq Effective Single-Channel\rq\rq, correspond to $\mathcal{A}$ of Eqs.\,(\ref{A_n_exact}), (\ref{A_n_asymptotic}) and (\ref{A_n_single}), respectively.
}
\label{fig:model}
\end{figure}

Finally, we show an example of the total survival probability, which includes contributions of multiple poles.
We adopt a model in which a bare state is coupled to two open channels with a zero-range interaction \cite{Braaten:2007nq} for simplicity.
Due to the zero-range interaction, peculiarities exist in the small-time behavior of a power of $t^{1/2}$ \cite{RamirezJimenez:2021pok}.
Here, we will not be concerned with this short-time behavior.
There are four poles in this model, and ${\cal G}$ is given by the Flatte form \cite{FLATTE1976224} as,
\begin{align}
  {\cal G}
   = \frac{1}{{\varepsilon} - m + i \gamma_1 \sqrt{\varepsilon} + i \gamma_2 \sqrt{\varepsilon-1}}=\sum_{n=1}^4\frac{r_n}{z-z_n}.
\end{align}
Fig.\,\ref{fig:model} shows the energy spectrum and the survival probability for the case $m = 1.1$, $\gamma_1=0.1$, $\gamma_2=0.9$.
The pole positions are $z_1=0.861-0.087i$, $z_2=0.707-0.913i$, $z_3=-z_1^\ast$ and $z_4=-z_2^\ast$.
For this parameter set, the energy spectrum and the survival probability of $t>0$ are dominated by a single pole, pole 1, which corresponds with the \lq\lq threshold cusp\rq\rq~pole of case B in the argument above.
\par
In conclusion, we have clarified the time dependence of unstable states in the context of two channels. 
We have shown that not only the energy spectrum but also the time dependence of unstable states are determined by the minimal distance in energy between the pole and the physical region, not by the imaginary part of the complex pole energy, in general.
This fact emerges most crucially in the time dependence of near-threshold unstable states, in particular, \lq\lq threshold cusp\rq\rq~states,
whose pole energy has a positive imaginary part but decays, not grow in time.
We have also shown that the decay of the \lq\lq threshold-cusp\rq\rq~is non-exponential in contrast to the resonance. 
Thus,  we have shown that the \lq\lq threshold-cusp\rq\rq~is a new class of unstable mode, which shows up only in coupled channels.
\par
In experimental nuclear or particle physics, it would be extremely difficult or almost impossible to observe the non-exponential decay of the \lq\lq threshold cusp\rq\rq~found in the present letter due to the characteristic time scale of the order of $100 \, {\rm MeV}^{-1} \sim 10^{-23} {\rm s}$.
Some materials might have an excitation mode with multiple decay channels, in which the relation between the associated pole and the thresholds of decay channels is similar to the \lq\lq threshold cusp\rq\rq~discussed in the present letter.
It would be exciting if such an excitation were found, given that non-exponential behavior at large times was found in an experiment measuring the luminescence decays of dissolved organic materials \cite{rothe2006violation}.
\par
O. M. would like to thank the members of the discussion meeting held in KEK Tokai campus: Yoshinori Akaishi, Akinobu Dote, Toru Harada, Yudai Ichikawa, Fuminori Sakuma, Shoji Shinmura, and Yasuhiro Yamaguchi.
K. Y. is grateful to iTHEMS at RIKEN and the director T. Hatsuda in particular for providing wonderful and stimulating working circumstances.
This work was supported by JSPS KAKENHI Grant Numbers JP22J15277, JP22KJ0982.
\bibliography{survival}
\end{document}